\def\eq{\ =\ }
\def\mns{\ -\ }
\def\pls{\ +\ }
\documentclass[12pt,a4paper,epsfig]{iopart}
\usepackage{epsfig}
\begin{document}
\title{Augmented  space recursion for partially disordered systems.}
\author{\bf Atisdipankar Chakrabarti and Abhijit Mookerjee}
\address { S. N. Bose National Centre for Basic Sciences. 
Block-JD, Sector-III, Kolkata-700098, India.}
\ead{adc@boson.bose.res.in}
\vskip -0.5cm
\ead{abhijit@boson.bose.res.in}
\begin{abstract}{Off-stoichiometric alloys exhibit partial disorder, in the sense
that only some of the sublattices of the stoichiometric ordered alloy become
disordered. This paper puts forward a generalization of the augmented space
recursion (ASR) introduced earlier by one of us (Mookerjee \etal 1997(*)) for systems with many atoms per unit cell.
In order to justify the convergence properties of ASR we have studied 
the convergence of various moments of local density of states and other
physical quantities like Fermi energy and  band energy. We have also looked
at the convergence of the magnetic moment of Ni, which is very
sensitive to numerical approximations  towards the $k$-space value
0.6 $\mu_{B}$ with the number of recursion steps prior to termination. }
\end{abstract}
\submitto{\JPCM}
\maketitle
\parindent 0pt

\section{Introduction}   

Binary alloys in stoichiometric compositions invariably exhibit ordered structures
at low temperatures. As we depart from perfect stoichiometric compositions, it is not
possible to populate the lattice in the given compositions so as to produce a perfectly ordered 
structure. Take for example, a B$_{75}$A$_{25}$ binary alloy on an fcc lattice. One of the
possible stable ordered phases is the L12 arrangement as shown in figure 1. An example of this is
Cu$_3$Au. In a cubic unit cell the corner is occupied by an A atom, while the three face centres by B atoms.
Since there are N corners and 3N face centres (N being the total number of unit cells in the solid) 
 and exactly as many A and B atoms, at this composition 
the L12 ordered arrangement exactly populates all the lattice sites. Ordered arrangement becomes 
impossible, say for B$_{70}$A$_{30}$. However, the following arrangement becomes possible :
since there are now 1.2N A atoms, N of them may occupy the N corners.
The 3N face centres may be occupied randomly by the 2.8N B atoms and 0.2N remaining A atoms. The original A sublattice
remains ordered, while the B sublattice becomes disordered. Since there are on the whole 3N face centres, the
occupation probability of the A and B atoms in this sublattice are 0.9\.{3} and 0.0\.{6} respectively. This arrangement
is quite different from the completely random alloy, where all sites are randomly occupied by either the
A or B atom with probabilities 0.3 and 0.7 respectively. It is also rather different from the {\sl partial disorder}
defined by  \cite{kudr}. In this communication we shall define {\sl partial disorder} in the
manner described above.
\vskip 1cm

\begin{figure}[b]                                                             
\centering                                                                         
\epsfxsize = 2.5 in \epsfysize = 4.5in \rotatebox{270}{\epsfbox{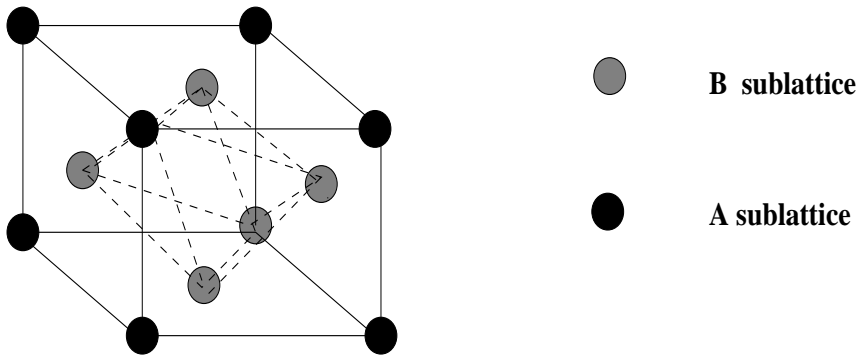}}
\caption{The L12 atomic arrangement for 75-25 AB binary alloys on a fcc lattice}
\label{fig1}                                                
\end{figure}                                    

The recursion method was introduced by  \cite{hhk,hk} as a convenient
and numerically efficient method for calculating Green functions and  realistic physical 
properties like the local density of states (LDOS), the Fermi energy and the band energy. 
The method comes to its own in situations where translation symmetry of the potential
in an effective one-electron Hamiltonian is lost. The Bloch Theorem is violated, and 
we cannot employ the standard reciprocal space techniques unless we carry out a homogeneity
restoring mean-field approximation (like the coherent potential approximation, CPA).
This can happen, for instance, at a surface or in a
random alloy. In particular, if the ion-cores sit on a topologically distorted network or at a
rough surface or interface, where the recursion method seems the only natural way of dealing
with electronic property calculations.  

There exists a large body of literature on the application of the recursion method to various
situations (\cite{v35}). For substitutional disorder, the recursion method has been
employed in tandem with the augmented space formalism (\cite{am1,am2,sdm1,sdm2,sm}) (augmented space recursion or ASR)
for dealing with random
alloys, taking into account local configuration fluctuations like chemical clustering, local lattice distortions due to size effects
and inhomogeneous randomness at a surface because of surface segregation. In spite of this 	
extensive body of literature, the recursion method does not seem to have gained wide acceptability
in the electronic structure community, and one often has had to face skepticism regarding its accuracy
and feasibility when applied with sophisticated electronic structure techniques to realistic systems.

In this communication we shall first present the ASR in detail for the most general case of systems
with many atoms per unit cell. Our subsequent aim will be to study partially disordered systems in 
alloys with off-stoichiometric compositions. 

We shall carry out  recursion calculations based on the tight-binding linearized muffin-tin orbitals (TB-LMTO)
method (\cite{oka1,oka2,oka3}) for various metals and alloys, to start with for perfectly ordered systems, compare them with standard
reciprocal space methods also based on the TB-LMTO and then carry out a thorough analysis of the 
convergence and accuracy of the technique. The aim will be to have a convincing handle on 
the possible errors in the method with a view to minimizing them within our tolerance limits.
Having convinced ourselves about the feasibility of the method, we shall first apply it to a toy model
for partial disorder before carrying on to realistic systems.
\section{Formalism}
\subsection{Generalized Augmented Space Recursion within the TB-LMTO formalism.}
In earlier works (\cite{sdm1,sdm2,sm}) it has been established that
for a disordered system the Augmented Space Theorem maps a disordered
Hamiltonian described in a Hilbert space $H$ onto an ordered Hamiltonian in
an enlarged space $\Psi$, where the space $\Psi$ is constructed by augmenting
the configuration space $\Phi$ of the random variables of the disordered 
Hamiltonian together with the Hilbert space $H$ of the disordered Hamiltonian
. Thus $\Psi=\Phi\otimes H$.  
The configuration average of the Green function reduces
to the evaluation of a particular matrix element of the resolvent of this
enlarged  Hamiltonian in the augmented space.
Hence, if one performs a recursion in the augmented space, one can obtain
the matrix element necessary to calculate the configuration average of
a Green function directly. The advantage of this method is that it does not
involve a single site approximation or the solution of any self-consistent
equation  (which is a prerequisite for the CPA or its generalizations).
Furthermore, one can treat both diagonal and off-diagonal disorder on an 
equal footing. In spite of its immense potential, the method could not
be implemented in a real alloy systems successfully earlier because of
the large rank of the augmented Hamiltonian. For a binary alloy with N
sites and with only {\it s}-orbitals it is   $N\times$2$^{N}$).
  However,  if we 
exploit the symmetry of the augmented space  (which arises due to the
homogeneity of the disorder) the rank of the irreducible part of the augmented space
in which the recursion is effectively carried out, is considerably   
reduced and recursion becomes tractable. Unlike other
methods like the embedded-cluster method, where the configuration averaging 
 is done by explicit 
averaging over all the distinct configurations of the embedded cluster  (144 for a fcc lattice
with first shell of neighbours), this generalized augmented-space recursion
technique yields, in a single recursion, the configuration average directly.
In a recent paper,  \cite{hay2} has rederived the augmented space technique
and related it to the idea of spatial ergodicity which connects volume averaging to
configuration averaging in the infinite solid limit. 

As has been mentioned earlier, since the recursion method needs a localized, 
short-ranged basis for its operation, one can implement augmented space
recursion in the framework of the TB-LMTO formalism. We now describe the 
methodology of generalized augmented space recursion in the framework of
the TB-LMTO.

The second order TB-LMTO Hamiltonian is written in terms of potential
parameters and screened structure matrix. For a random binary alloy
A$_{x}$B$_{y}$,  the LMTO Hamiltonian in the most localized representation
is given by, 

\begin{equation}
{\cal H}^{(2)} = {E}_{\nu} + h -h{o}h
\end{equation}
where, \\

\begin{equation}\fl  h  =  \sum_{RL\alpha} \left(\tilde{{\bf C}}_{RL\alpha} - \tilde{{\bf E}}_{R L \alpha}\right) \enskip {\cal P}_{RL\alpha} + 
\sum_{RL\alpha}\sum_{R'L'\alpha '} \tilde{\mathbf{\Delta}}_{RL\alpha}^{1/2}\ S_{RL\alpha, R'L'\alpha '}\ 
\tilde{\mathbf{\Delta}}_{R'L'\alpha '}^{1/2} \enskip {\cal T}_{RL\alpha, R'L'\alpha '} \end{equation}
 C, o and $\Delta$ are potential parameters of the TB-LMTO method, these are diagonal matrices in the angular momentum indices. Also $o^{-1}$ has the dimension of energy and is a measure of the energy window around
${\tilde E}$ around which the approximate Hamiltonian ${\cal H}^{2}$ is reliable.

\begin{eqnarray*}
\tilde{{\bf C}}_{RL\alpha} & = & C^{A}_{L}\ n_{R}^{\alpha} + C^{B}_{L}\ \left  ( 1-n_{R}^{\alpha} \right) 
 =  C^{B}_{L} + \delta C_{L}\ n_{R}^{\alpha} \nonumber\\
\delta C_{L} & = & C^{A}_{L} - C^{B}_{L} \nonumber\\
\tilde{\mathbf{\Delta}}_{RL\alpha}^{1/2} & = & \left  ( \Delta^{A}_L \right)^{1/2} n_{R}^{\alpha}
+ \left  (\Delta^{B}_L \right)^{1/2} \left  (1-n_{R}^{\alpha} \right) 
= \left  ( \Delta^{B}_L \right)^{1/2} + \delta \Delta_{L}^{1/2}\ n_{R}^{\alpha} \nonumber\\
\delta\Delta_{L}^{1/2} & = & \left  (\Delta^{A}_L \right)^{1/2} - \left(\Delta^{B}_L \right)^{1/2} \nonumber \\
\tilde{{\bf o}}_{RL\alpha} & = & o^{A}_{L}\ n_{R}^{\alpha} + o^{B}_{L}\ \left( 1-n_{R}^{\alpha} \right) 
 =  o^{B}_{L} + \delta o_{L}\ n_{R}^{\alpha} \nonumber \\
\delta o_{L} & = & o^{A}_{L} - o^{B}_{L} \nonumber \\
\tilde{{\bf E}}_{\nu RL\alpha}& = & E^{A}_{\nu L\alpha}n^{\alpha}_{R}+E^{B}_{\nu L\alpha}
\left (1-n^{\alpha}_{R}\right) 
= E^{B}_{\nu L\alpha}-\delta E_{\nu L\alpha}\ n_{R}^{\alpha} \nonumber \\
\delta E_{\nu L\alpha} & = & E^{A}_{\nu L\alpha}-E^{B}_{\nu L\alpha} \nonumber
\end{eqnarray*}

\noindent $R$ denotes a {\sl cell position} label associated with a TB-LMTO basis and $L$ = ($\ell m m_s$) is the
composite angular momentum index. $\alpha$ denotes an atom in the $R$-th cell whose position is $R+\xi^\alpha$.
$n_{R}^{\alpha}$ is the site-occupation variable which takes values 0 or 1
depending upon whether site $\alpha$ in the $R$-th cell is occupied by an A or a B atom. For {\sl partial disorder}
this is a random variable whose probability density depends upon which sublattice it belongs to, hence the
label $\alpha$ associated with it. The structure
matrix S$_{RL\alpha, R'L'\alpha '}$ is non-random in case of substitutional alloys with
negligible size mismatch. Now one can obtain the full ${\cal H}^{2}$ by
inserting $h$ in expression (1).

\noindent ${\cal P}_{RL\alpha}$ and ${\cal T}_{RL\alpha, R'L'\alpha '}$ are the projection and transfer
operators in Hilbert space $H$ spanned by tight-binding basis $\left\{\vert RL\alpha \rangle \right\}$.

\begin{eqnarray}
{\cal P}_{RL\alpha} & = & \vert RL\alpha \rangle \langle RL\alpha \vert \nonumber \\
{\cal T}_{RL\alpha, R'L'\alpha '} & = & \vert RL\alpha \rangle \langle R'L'\alpha ' \vert 
\end{eqnarray}

The expanded Hamiltonian $\hat{\cal H}$ in the  augmented space
is constructed by replacing the random site occupation variable $n_{R}^\alpha$ by
its corresponding operator representation ${\cal M}_{R}^{\alpha}$ in configuration
space. 

\begin{eqnarray}
\fl \hat{\mathbf{h}}  =  \sum_{RL\alpha} \left(\left(C^{B}_{L}-E^{B}_{\nu L}\right)\  \hat{{\cal I}} +
\left(\delta C_{L}-\delta E_{\nu L}\right)\ {\cal M}_{R}^{\alpha} \right) \otimes {\cal P}_{RL\alpha} +\ldots \nonumber\\
\fl  +  \sum_{RL\alpha} \sum_{R'L'\alpha} \left(   (\Delta^{B}_{L})^{1/2}\  \hat
{{\cal I}} + \delta \Delta^{1/2}_{L}\  {\cal M}_{R}^\alpha \right) S^{\alpha}_{RL\alpha, R'L'\alpha '} 
\left(   (\Delta^{B}_{L'})^{1/2}\  \hat{{\cal I}}  +
\delta {\Delta}^{1/2}_{L'} {\cal M}_{R'}^\alpha \right) \otimes {\cal T}_{RL\alpha,R'L'\alpha '}\nonumber \\
\end{eqnarray}

\noindent where, ${\cal M}_{R}^\alpha$ is given by 

\[ {\cal M}_R^\alpha \ =\  x_{A}^{\alpha} {\cal P}_{R\alpha}^{\uparrow} + x_{B}^{\alpha} {\cal P}_{R\alpha}^{\downarrow}+ \sqrt{(x_A^\alpha x_B^\alpha)}
\left( {\cal T}_{R\alpha}^{\uparrow\downarrow}+{\cal T}_{R\alpha}^{\downarrow\uparrow}\right)\]
Now inserting equation(4) in equation(1) one can obtain the full Hamiltonian $\hat{\cal H}$
This Hamiltonian is now an operator in a much enlarged space $\Psi = \Phi \otimes{\cal H}$
 (the augmented space). The Hilbert space $\Psi$ is 
spanned by the  basis set $\left\{ \vert R, L, \alpha, \{C\}\rangle \right\} $, 
where $\{C\}$ is a pattern ($\uparrow\uparrow\downarrow\uparrow\ldots\}$) spanning the
configuration space and described in terms of the {\sl cardinality sequence} $\{C\}$ of
positions of the $\downarrow$ in the pattern. The enlarged Hamiltonian does not involve any random variables but
incorporates within itself the full statistical information about the random occupation
variables.

In the presence of
off-diagonal disorder,  which is invariably present in the form of the
TB-LMTO Hamiltonian,  even the reduction of the rank of the invariant subspace
on which recursion acts, using the symmetries of the augmented space, 
does not allow us to sample as many configuration
states as we would like. This is because,  as recursion proceeds,  the
number of configuration states sampled at the n-th step of recursion
becomes unmanageably large when n $>$ 5   (for example). To do away with
this problem,  the working equations are transformed so as to put the
Hamiltonian for the recursion in an effective diagonal disorder form.
This allows one to sample further shells in augmented-space and to confirm
the shell-convergence of the recursion.

To do this,  we first suppress all the indeces and write the expression for the resolvent as
follows :

\[
\left ( E - {\cal H }^{(2)} \right) ^{-1}   =  \left ( {E} - {\mathbf{\tilde C}} - 
{\mathbf{\tilde \Delta}}^{1/2} 
S{\mathbf{\tilde \Delta}}^{1/2}+{h}{\mathbf{\tilde o}}{h} \right) ^{-1} \]
\[
\fl  \ =\ {\mathbf{\tilde \Delta}}^{1/2}  \left[\frac{E -{\mathbf{\tilde C}}}
{{\mathbf{\tilde \Delta}}} -S+\left (
\frac{{\mathbf{\tilde C}}-{\mathbf{\tilde E}}_{\nu}}{{\mathbf{\tilde \Delta}}}+S \right ) 
 \left ({\mathbf{\tilde \Delta}}^{1/2}{\mathbf{\tilde o}}{\mathbf{\tilde \Delta}}^{1/2}\right )
\left (\frac {{\mathbf{\tilde C}}-{\mathbf{\tilde E}}_{\nu}}{{\mathbf{\tilde \Delta}}}+S \right )\right ]^{-1} {\mathbf{\tilde \Delta}}^{1/2} 
\]

Using the augmented space theorem,  we can write the expression of 
configuration averaged Green function as, 

\begin{eqnarray*}
\fl \ll G_{RL\alpha,RL\alpha}  (E) \gg & = & \langle R,L,\alpha, \{\emptyset\} \vert   
\left   (E\hat{I}-\hat{\cal{H}} \right)^{-1} \vert R,L,\alpha, \{\emptyset\} \rangle
\end{eqnarray*}

where, 
\[  A^{\alpha}_L\left(\Delta^{-1/2}\right)\vert R, L, \alpha \otimes \{\emptyset\}\rangle
+ F^{\alpha}_L\left(\Delta^{-1/2}\right)\vert R,L,\alpha \otimes \{R\}\rangle =\vert 1\}\] 

We define $\left [ A^{\alpha}_L\left (1/\Delta\right )\right]^{1/2}\vert 1\}$ as the normalized
initial recursion vector $\vert 1\rangle$.

Finally,  we arrive at the convenient expression  :

\[
\fl \ll G_{RL\alpha,RL\alpha}  (E) \gg  =  \langle 1\vert\left [\hat{E}-\hat{A}+\hat{B}+\hat{F}-\hat{S}+\left(\hat{J}+\hat{S}\right)
\hat{o}\left(\hat{J}+\hat{S}\right)\right]^{-1}\vert 1\rangle 
\]

\noindent where,

\begin{eqnarray}
\hat{A} &=& \sum_{R,L,\alpha}\left\{\frac{A^{\alpha}_L\left (C/\Delta\right)}{A^{\alpha}_L\left(1/\Delta\right)}\right \}{\cal P}_{R,\alpha}
\otimes {\cal P}_L \otimes {\cal I} \nonumber \\
\hat{B} &=& \sum_{R,L,\alpha} \left\{\frac{B^{\alpha}_L\left (\left (E-C \right)/\Delta\right )}{A^{\alpha}_L \left(1/\Delta \right )} \right \} {\cal P}_{R,\alpha}\otimes {\cal P}_L\otimes{\cal P}^{\downarrow}_{R\alpha} \nonumber \\
\hat{F} &=&\sum_{R,L,\alpha} \left\{\frac{F^{\alpha}_L\left (\left (E-C\right )/\Delta \right )}{A^{\alpha}_L\left (1/\Delta \right )}\right \} 
{\cal P}_{R,\alpha}\otimes {\cal P}_{L} \otimes 
\left( {\cal T}_{R\alpha}^{\uparrow\downarrow} +
 {\cal T}_{R\alpha}^{\downarrow\uparrow} \right)\nonumber 
\end{eqnarray}
\[ \fl 
\hat{S}   =   \sum_{RL\alpha}\sum_{R'L'\alpha '}\left\{A^{\alpha}_L 
\left( 1/\Delta \right)^{-1/2} \right \} S_{RL\alpha, R'L'\alpha '}
 \left \{A^{\alpha}_L\left(1/\Delta\right)^{-1/2} \right\} 
{\cal T}_{R\alpha,R'\alpha '}\otimes {\cal T}_{LL'} 
\otimes {\cal I} 
\]

\noindent  here, $\hat{J}=\hat{J}_{A}+\hat{J}_{B}+\hat{J}_{F}$ and $\hat{o}=\hat{o}_{A}+\hat{o}_{B}+\hat{o}_{F}$, where,

\begin{eqnarray}
\hat{J}_{A} &=&  \sum_{R,L,\alpha}\left\{\frac{A^{\alpha}_L\left(\left (C-E_{\nu}\right )/\Delta \right )}
{A^{\alpha}_L\left (1/\Delta \right )}\right \} {\cal P}_{R,\alpha}\otimes {\cal P}_L\otimes {\cal I}\nonumber \\
\hat{J}_{B} &=& \sum_{R,L,\alpha} \left\{\frac{B^{\alpha}_L\left(\left (C-E_{\nu}\right )/\Delta\right )}{A^{\alpha}_L\left (1/\Delta \right )}\right\}{\cal P}_{R,\alpha}\otimes{\cal P}_L\otimes{\cal P}^{\downarrow}_{R\alpha} \nonumber \\ 
\hat{J}_{F} &=& \sum_{R,L,\alpha} \left \{\frac{F^{\alpha}_L\left (\left (C-E_{\nu}\right )/\Delta \right )}{A^{\alpha}_L\left (1/\Delta \right )}\right \}{\cal P}_{R,\alpha} \otimes {\cal P}_{L}\otimes \left  ( {\cal T}_{R\alpha}^{\uparrow\downarrow} + {\cal T}_{R\alpha}^{\downarrow\uparrow} \right) \nonumber \\
\hat{o}_{A} & = & \sum_{R,L,\alpha}\left \{ A^{\alpha}_L\left({\tilde o}\right)A^{\alpha}_L\left (1/\Delta\right)\right \}{\cal P}_{R,\alpha}\otimes
{\cal P}_L\otimes {\cal I} \nonumber \\
\hat{o}_{B} &=& \sum_{R,L,\alpha} \left \{ B^{\alpha}_L\left({\tilde o}\right)A^{\alpha}_L\left (1/\Delta\right)\right \}{\cal P}_{R,\alpha}\otimes{\cal P}_L\otimes{\cal P}^{\downarrow}_{R\alpha} \nonumber \\
\hat{o}_{F} &=& \sum_{R,L,\alpha} \left \{ F^{\alpha}_L\left({\tilde o}\right)A^{\alpha}_L\left (1/\Delta\right)\right \}{\cal P}_{R,\alpha} \otimes{\cal P}_{L} \left  ( {\cal T}_{R\alpha}^{\uparrow\downarrow} + {\cal T}_{R\alpha}^{\downarrow\uparrow} \right) \nonumber 
\end{eqnarray}
\noindent where,
\begin{eqnarray}
A^\alpha_L  (Z) & = & x_A^\alpha\ Z^{A}_L + x_B^\alpha\ Z^{B}_L \nonumber\\
B^\alpha_L  (Z) & = & \left  ( x_B^\alpha -x_A^\alpha \right)\ \left  (Z^{A}_L-Z^{B}_L \right) \nonumber\\
F^\alpha_L  (Z) & = & \sqrt{x_A^\alpha x_B^\alpha}\ \left  (Z^{A}_L-Z^{B}_L \right) \nonumber
\end{eqnarray}
Z is any single site parameter.

Though the computational burden is considerably reduced due to diagonal
formulation,  the recursion now becomes energy dependent as is clear from
the expressions of $\hat B$ and $\hat F$ above. The old formalism was free of this constraint.
This energy dependence makes the recursion technique rather unsuitable
because now we have to carry out one recursion per energy point of interest.
This problem is tackled using {\it seed recursion technique} (\cite{gdm}). The idea is
to choose a few seed points across the energy spectrum uniformly,  carry out
recursion on those points and then fit the  coefficients of recursion
throughout the whole spectrum. This way one can save huge computation time.
 For example, if one is interested in an energy spectrum of 100 points, 
in the bare diagonal formulation recursion has to be carried out at all the 100
points but in the seed recursion technique one needs to perform recursions
only at 15-20 points. The whole idea stems from the fact that in most of the
cases of binary alloys,  it is seen that the recursion coefficients 
$\alpha_{n}$ and $\beta_{n}$ vary quite weakly across the energy spectrum.
So, one can easily pick up a few of them and fit throughout the whole range
of energy by a suitable function. 

\subsection{Convergence of the recursion method}
Before we apply the above methodology to study realistic alloy systems,
we must first ensure that the technique is stable and convergent.
We shall  illustrate the nature of convergence
of various calculated physical quantities which are basic for any
electronic structure calculation. This study of convergence is essential
because otherwise we will not be able to set the cut-offs of various
parameters like number of shells in real and configuration spaces, 
recursion levels and so on, for further calculations and ensure that
our numerical results are within our tolerance window.

\begin{figure}[t]
\centering                                                                         
\epsfxsize = 5.5 in \epsfysize = 5in \rotatebox{-90}{\epsfbox{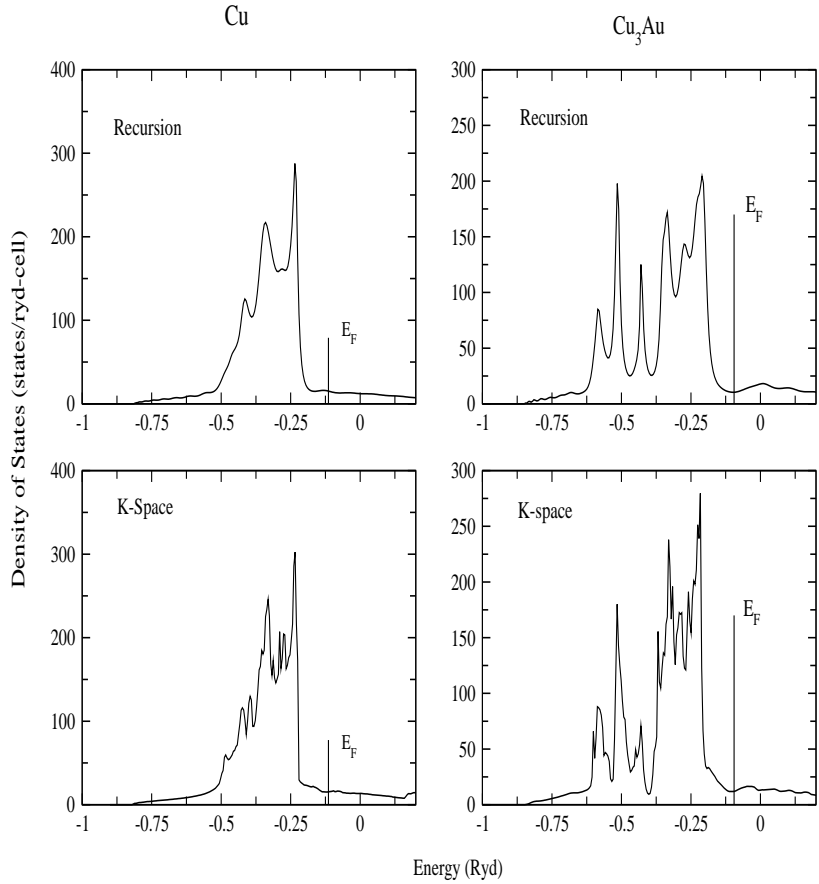}}
\caption{The Density of States for Cu and Cu$_3$Au using (top)  the recursion method
and (bottom) the k-space method}
\label{fig2}                                                
\end{figure}

\begin{figure}[t]
\centering                                                                         
\epsfxsize = 5.5 in \epsfysize = 5in \rotatebox{-90}{\epsfbox{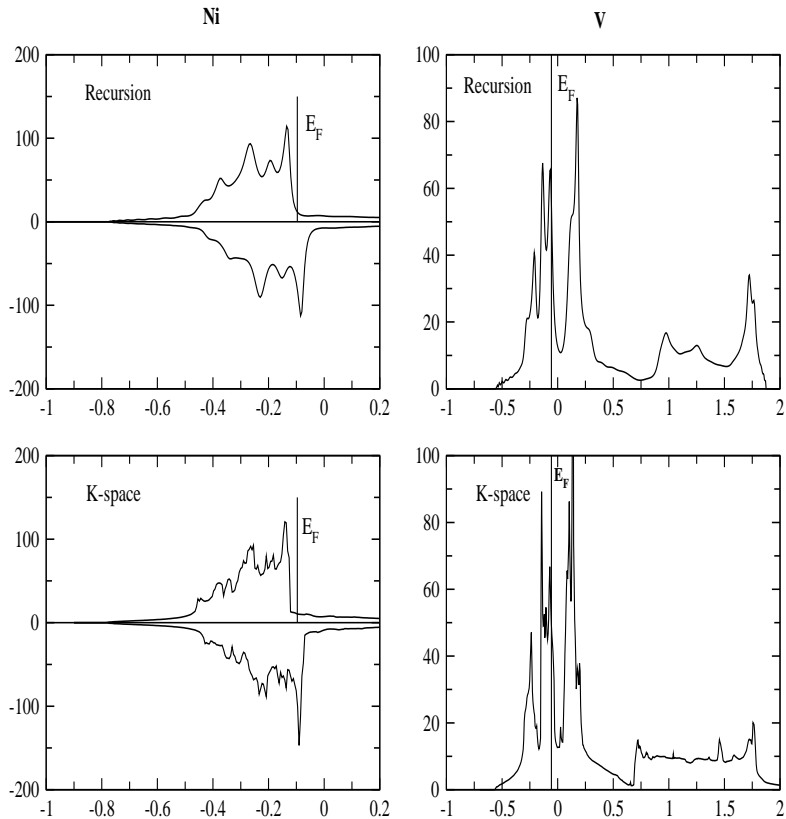}}
\caption{The Density of States for Ni and V using (top)  the recursion method
and (bottom) the k-space method}
\label{fig3}                                                
\end{figure}

Before we set out to analyze the errors in the method, let us first compare the
results for the density of states for Cu, Cu$_3$Au, V and Ni as obtained from
a forty-step recursion carried out on an exact twenty shell real-space map.
The choice of the systems is deliberate. Cu is a full $d$-shell noble metal,
V is a half-filled  metal in the lower end of the transition series while Ni is
a magnetic transition metal. Cu$_3$Au is an ordered alloy which is stable in the L12
arrangement on a fcc lattice.  Figures \ref{fig2} and \ref{fig3} show this comparison.
\vskip 1cm

For $Cu$ and $Cu_3Au$ most of the detailed features of the density of states calculated
from the recursion calculations match with the k-space results. As expected, the k-space
results show sharper structures. The Fermi energies in both cases match to within a
few hundredths of a rydberg. Both the majority and minority band density of states for
ferromagnetic $Ni$ are reproduced excellently well in the recursion calculations.
The agreement is equally good for $V$ upto the Fermi energy.
The unoccupied part of the band shows disagreement. However, it must be understood that
the recursion is done with the second order Hamiltonian in the most localized representation
of the TB-LMTO, while the k-space calculations are done with the orthogonal representation.
A higher order recursion calculation should improve  the unoccupied part of the
band considerably.

How good are our results ?
When we talk of convergence of the recursion method,  we have to be careful
in stating precisely what we mean. Finite space approximants to Green
functions do not converge for real energy values. This problem arises in
every computational method,  as noted by \cite{v35} . The problem
definitely arises in the usual k-space integration techniques,  where
methods using either complex energies or complex k-s have been attempted.
The cause of this non-convergence is that an arbitrary small perturbation, 
like adding a single atom to a large but finite system,  can shift all
eigenvalues of the system. This causes an infinite change in the Green
function near its corresponding poles. Thus,  the precise meaning of the
convergence of the recursion should imply rather the convergence of
physical quantities built out of it. Most physical quantities are averages
over the spectrum of the type  :

\[ F  (E)\ =\ \int_{E_0}^{E_F} f  (E')\; n  (E') dE' \]

$E_0$ is the lower band edge, and $f(E)$ is any smooth, well-behaved function of $E$. 
It is the convergence of these quantities which will decide whether the
recursion is convergent or not. For example,  the Fermi energy is defined by

\[  \int_{-E_0}^{E_{F}} n  (E')dE' \eq n_{e} \]

\noindent where $n_{e}$ is the total number of electrons.  While the band energy
is

\[ U \eq \int_{-E_0}^{E_{F}} E'\; n  (E') dE' \]

\noindent We shall study, in general, the convergence of indefinite integrals of the 
kind 

\[ M_{k}  (E)\eq\int_{-\infty}^{E}   (E')^{k}\; n  (E') dE' \]

\noindent The integrand $E^{'^{k}} \;$
is monotonic and well behaved within the integration range. A measure of the  root-mean-square error in the moments is

 \begin{equation} \fl \Delta_k  \eq  \left\{ \frac{1}{E_U - E_0}  \left(\int_{E_0}^{E_U} \left(\delta M_k (E)\right)^2 dE\mns \left(\int_{E_0}^{E_U} \delta M_k (E) dE\right)^2\right)\right\}^{1/2} \label{err}\end{equation}

Errors can arise in the recursion procedure because of two distinct
sources  :    (i) the error that arises because we carry out a finite number
of recursion steps and then terminate the continued fraction using one of
the available terminators ;   (ii) the error that arises because we choose a
large but finite part of the nearest-neighbour map and ignore the part of
the augmented space very {\sl far} from the starting state. Haydock has justified
both these approximations by stating that (i) if the terminator is so chosen so as
to reflect the asymptotic behaviour of the continued fraction, errors should be small,
and (ii) since we can write expressions for the continued fraction in terms of self-avoiding
{\sl walks} in augmented space, long walks are dominated by those that wind round the
starting state and do not go far away from it.

We shall first carry out a simple error analysis of the continued fraction expression for 
the Green function because of errors created on the continued fraction coefficients. 

The recursion is a two-term recurrence relation. We may therefore generate from this,  a pair of
linearly independent  set of polynomials through the relations  : 

\[
b_{n+1}X_{n+1}  (E)  \eq    (E\mns a_{n})X_{n}  (E) \mns b_{n}X_{n-1} 
\]

\noindent where,  $X_{n}(E)$ is either $P_{n}(E)$ or $Q_{n}(E)$ according to the initial conditions  : \\

\begin{center}
\begin{tabular}{cccccc}
$P_{1}  (E)$ &\eq & 1 & $P_{2}  (E)$ &\eq  & $  (E\mns a_{1})/b_{2}$ \\
$Q_{1}  (E)$ &\eq & 0 & $Q_{2}  (E)$ & \eq & 1 \\
\end{tabular}
\end{center}
The approximated Green function in terms of the terminator $T(E)$ is given by  : 

\[ G  (E) \eq \frac{Q_{N+1}  (E)\mns b_{N}Q_{N}  (E)T  (E)}{b_{1}\left[ P_{N+1}  (E)\mns
b_{N}P_{N}T  (E) \right]} \]

The terminator determines entirely  the essential singularities of the
the spectrum.  \cite{v35} showed that a finite composition of
fractional linear transformations like the one above can at most add a finite
number of poles to the spectrum. The essential singularities of the exact $G(E)$ and
$T(E)$ coincide. The fractional linear transformation  redistributes
the spectral weights over the spectrum.

Let us now assume that we make errors $\{ \delta a_{n},  \delta b_{n} \}$ in the corresponding continued fraction
coefficients.  If we now start generating the orthogonal polynomials,  starting from the exact initial conditions, 
but with the errors in the continued fraction coefficients,  we shall obtain a pair of sets $\{ \tilde{P}_{n} \}$
and $\{ \tilde{Q}_{n} \}$ . In general we shall have, 

\[ {\mathaccent 126 P}_{n}  (E) \eq \left  (1+ A_{n}  (E)\right) P_{n}  (E) + B_{n}  (E) Q_{n}  (E) \]

If we substitute this back into the recurrence relation and keep only the first order terms in the errors, 

\begin{figure}[t]                                                               
\centering                                                                         
\epsfxsize = 5.5 in \epsfysize = 5in \rotatebox{-90}{\epsfbox{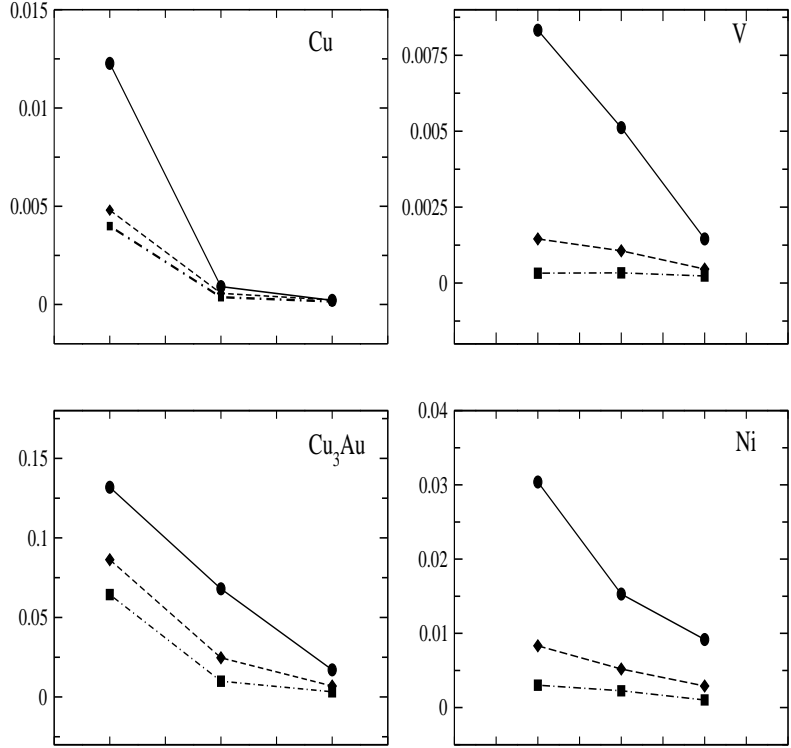}}
\caption{The root mean square errors in the first (full lines), second (dashed lines) and third (dashed dotted
lines) moments of the density of states as functions of the recursion steps}
\label{fig4}                                                
\end{figure}

\begin{eqnarray*}
A_{n}  (E) & \eq &  \left\{ \delta a_{n} \left[ P_{n+1}  (E) Q_{n+1}  (E) \right]\right.\nonumber\\ 
               & & \left. +
\delta b_{n} \left[ P_{n}  (E)Q_{n+1}  (E)+ P_{n+1}  (E)Q_{n}  (E)\right]\right\}/b_{1}\nonumber\\
B_{n}  (E) & \eq & \left\{ - \delta a_{n} P_{n+1}  (E)^{2} 
\mns \delta b_{n} \left[
2\; P_{n}  (E)P_{n+1}  (E)\right]\right\}/b_{1} \nonumber\\
\end{eqnarray*}

Using the above and the expression for the local density of states 
,  we find that the first order relative error produced in the local density of states

\[
\frac{\delta n  (E)}{n  (E)}\eq -2\left[ \left\{ \sum_{n=1}^{\infty}\; A_{n}  (E)\right\}
 + b_{1}R  (E)\left\{\sum_{n=0}^{\infty} \;B_{n}  (E)\right\} \right]
\]

\noindent where $R(E)$ \eq ${\cal R}$e  $G(E)$.  If we define the weighted Hilbert transforms of
$P_{n}  (E)$ as the so-called associated functions  :  

\[ {\cal Q}_{n}  (E) \eq {\cal R}e \left\{ \int_{-\infty}^{\infty} \frac{P_{n}  (E')
n  (E')}{E-E'} dE' \right\} \]

These associated functions are also solutions of the three-term recursion. They
are not polynomials,  but are nevertheless orthogonal to the set $P_{n}(E)$.

In terms of these,  the error in the density of states is  : 

\begin{eqnarray}
\frac{\delta n  (E)}{n  (E)}& \eq & \frac{2}{b_{1}} \left\{ \sum_{n=1}^{\infty} \left[ \delta a_{n}
P_{n+1}  (E){\cal Q}_{n+1}  (E)
 + 2\;\delta b_{n+1} P_{n+1}{\cal Q}_{n+2}
 \right] \right\}\nonumber\\
\end{eqnarray}

If the first N continued fraction coefficients are exact, that is in case we carry out our
recursion on a N-shell neighbour map upto N steps and then terminate, the  error in the 
various moment functions are  : 

\begin{equation}
\delta M_{k}  (E) \eq  + 
\sum_{n=N}^{\infty} \left\{\delta a_{n} A_{n}^{  (k)}  (E)
+ \delta b_{n+1} B_{n}^{  (k)}  (E) \right\}
\end{equation}

Where, 
\begin{figure}[t]
\centering                                                                         
\epsfxsize = 5.5 in \epsfysize = 5in \rotatebox{-90}{\epsfbox{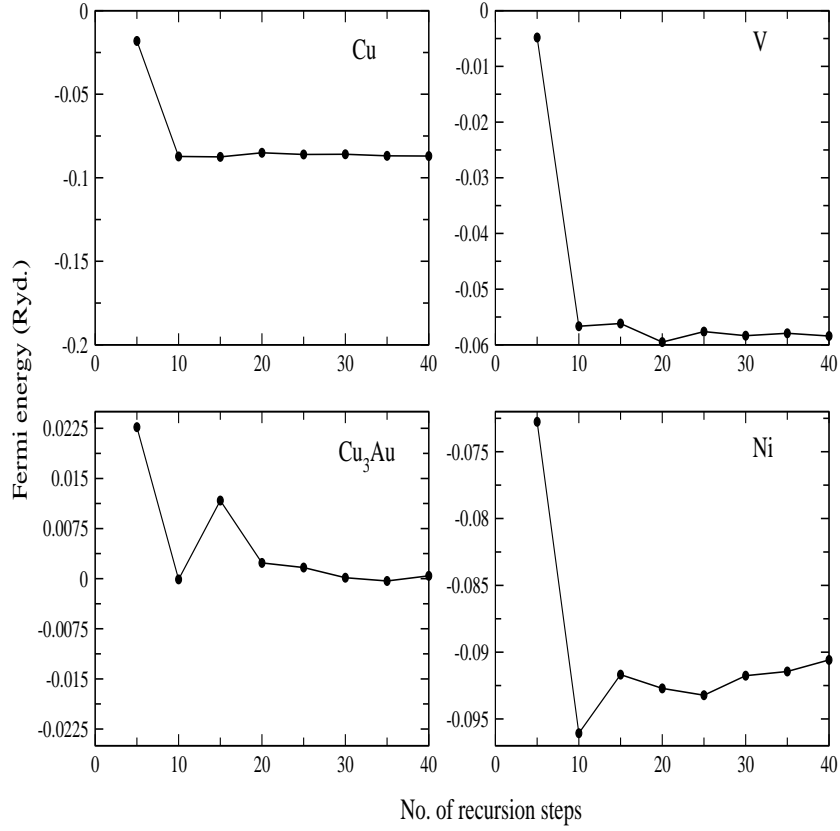}}
\caption{The convergence with recursion step of the Fermi energies of Cu, Cu$-3$Au, Ni and V}
\label{fig5}                                                
\end{figure}

\begin{eqnarray*}
A_{n}^{  (k)}  (E) & \eq & \frac{2}{b_{1}}\int_{-\infty}^{E} P_{n+1}  (E') {\cal Q}_{n+1}  (E')   (E')^{k} n  (E')dE'
\nonumber \\
B_{n}^{  (k)}  (E) & \eq & \frac{4}{b_{1}}\int_{-\infty}^{E} P_{n+1}  (E'){\cal Q}_{n+1}  (E')   (E')^{k} 
n  (E') dE'\nonumber \\
\end{eqnarray*}

\begin{figure}
\centering                                                                         
\epsfxsize = 5.5 in \epsfysize = 5in \rotatebox{-90}{\epsfbox{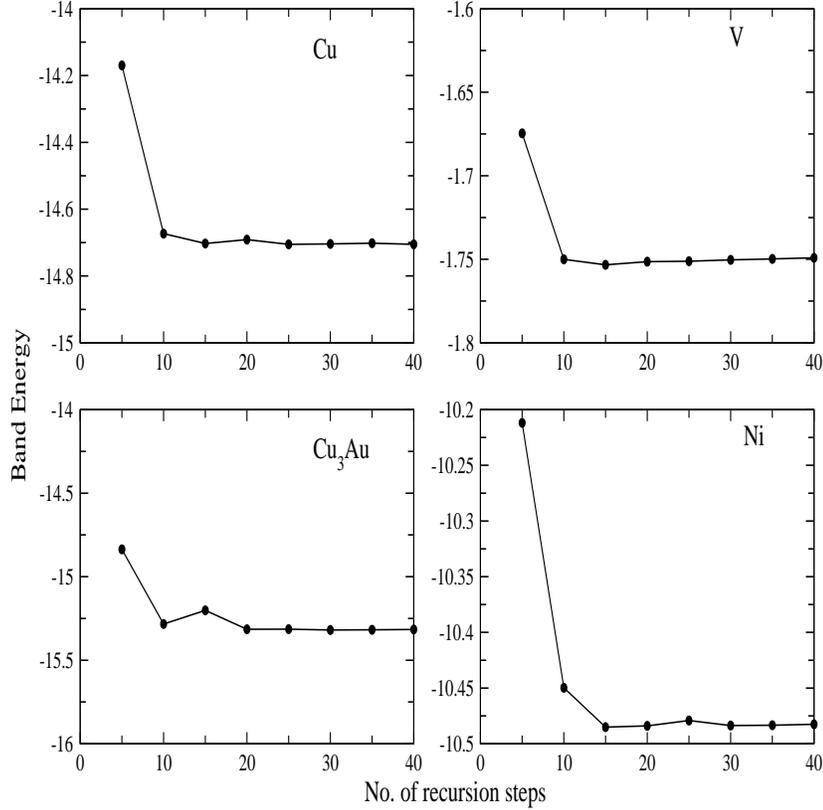}}
\caption{The convergence with recursion step of the band energies of Cu,Cu$_3$Au, Ni and V}
\label{fig6}                                                
\end{figure}

From this expression and equation(6) we can obtain an expression for the overall error in the moments.
Numerical results for the errors in the moment functions are shown in figure \ref{fig4}. The convergence in Fermi and band energies with number of recursion steps are shown in figures \ref{fig5} and \ref{fig6}.

\begin{figure}
\centering                                                                         
\epsfxsize = 4.5 in \epsfysize = 4in \rotatebox{-90}{\epsfbox{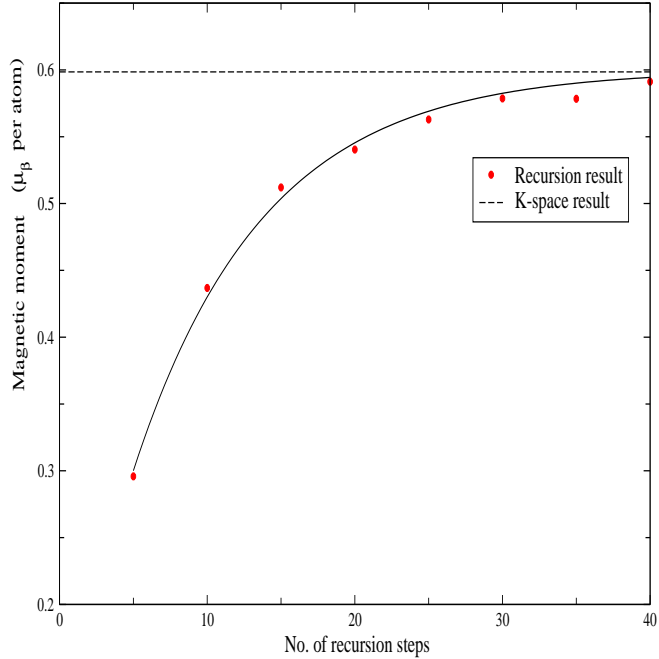}}
\caption{The convergence with recursion step of the magnetic moment of Ni}
\label{fig7}                                                
\end{figure}

These are calculated on a 20 shell neighbour map and upto 40 steps in recursion. The figures clearly illustrate the
convergence of the procedure with increasing recursion steps, as surmised by Haydock. As expected the higher moments
converge more rapidly. The rapid convergence of higher moments is at the basis of the reproduction of most
of the density of states shape in an approximate recursion procedure. It should be noted, however, that we carry out recursion beyond 40 steps, very soon the procedure becomes unstable. This instability arises due to two
sources : (i) we have carried out recursion on a 20 shell neighbour-map. As we go beyond 20 steps of recursions,
finite size effects begin to show up. Numerically till 40 steps these errors are tolerable. Beyond 45 they lead to
instability. We can control this by increasing the size of the neighbour-map. (ii) Numerical cumulative errors
lead to the loss of orthogonality in the recursion generated basis. This can be controlled by deliberately orthogonalizing after every 20 steps or so. 
In the presence of disorder, the sharp structures in the density of states are smoothened by finite life-time effects.The convergence of the higher moments is even more rapid. The above analysis in  ordered systems is therefore in a {\sl
worst scenario} situation.

Figure 7 shows the convergence of the magnetic moment of Ni as a function of the
number of recursion steps towards the value 0.6 $\mu B$ obtained from $k$-space
techniques. Magnetic moment is a physical quantity which is very sensitive to the
errors in numerical approximations. The figure shows that within 35-40 recursion
steps on a 20 shell real-space map, the magnetic moment has converged within our
error tolerance.

It is evident from the above analysis that the size of the neighbour map and the number of recursion steps required
can vary from system to system. For every situation, we have to carry out this error analysis before we can rely
on our numerical results with a degree of satisfaction.

\section{Calculations on a  model system}

\begin{figure}[t]
\centering                                                                         
\epsfxsize = 2 in \epsfysize = 2in \rotatebox{-90}{\epsfbox{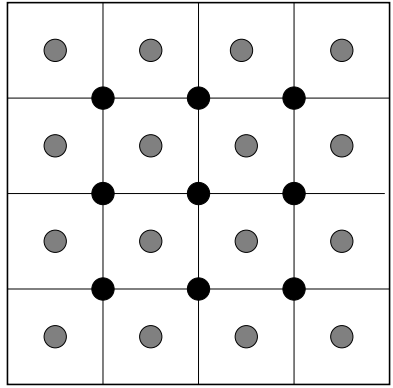}}
\caption{Ordered atomic arrangement on a square lattice for a A$_{50}$B$_{50}$ alloys}
\label{toy1}                                                
\end{figure}

\begin{figure}[t]
\centering                                                                         
\epsfxsize = 2.5 in \epsfysize = 2.0in \rotatebox{-90}{\epsfbox{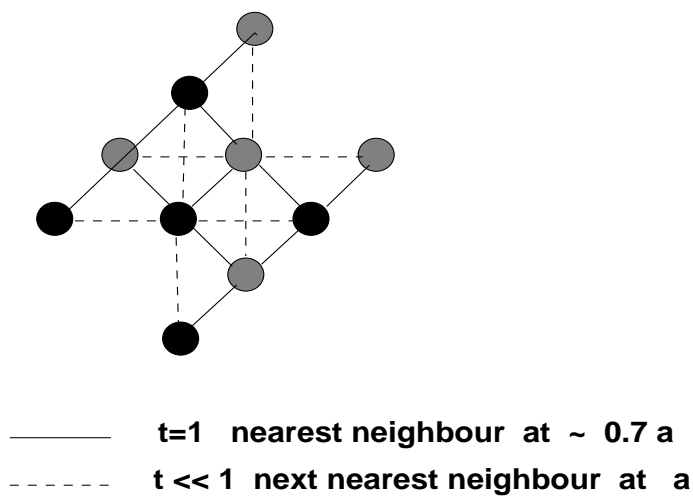}}
\caption{Nearest neighbour overlaps for a central cell in a square lattice with lattice
constant a units}
\label{toy2}                                                
\end{figure}

Before we carry out calculations on a real alloy, let us first apply our formalism to a toy model in order to
understand the effects of partial disordering. We shall consider a 50-50 AB alloy ordered first on a square
lattice as shown in figure \ref{toy1} and only $s$-states on this atomic arrangement. The Hamiltonian in a  tight-binding basis set is then,

\[ H \eq \sum_{i} \epsilon_{i\alpha,i\alpha'}\ {\cal P}_{i\alpha,i\alpha'} \pls \sum_{\{ij\}} t_{i\alpha,j\alpha '}\ {\cal T}_{i\alpha,j\alpha '} \]

where $\{ij\}$ denotes that i and j are nearest neighbour cells on the lattice, $\alpha$ is 1 or 2 according to
whether we a referring to the corner atoms or central atoms in a square unit cell (dark and light atoms 
in figure \ref{toy1}). If we consider only the nearest neighbour overlaps at a distance  $a/\sqrt{2}$ 
where $a$ is the square lattice constant, the diagonal and off-diagonal terms of the Hamiltonian are  
(referring to figure \ref{toy2})

\begin{eqnarray*}
 \epsilon_{i\alpha,j\alpha '} &\eq &\left(  \begin{array}{cc}
                                      \epsilon & t\\ t & \epsilon 
                                      \end{array} \right) \\
 t_{i\alpha,j\alpha '} &\eq &\left(  \begin{array}{cc}
                                      0 & 0\\ t & 0 
                                      \end{array} \right) \hspace{1cm}\mbox{$r_i$-$r_j$ are
in the  (10) and (01) directions }\\
 t_{i\alpha,j\alpha '} &\eq &\left(  \begin{array}{cc}
                                      0 & t\\ 0 & 0 
                                      \end{array} \right) \hspace{1cm}\mbox{$r_i$-$r_j$ are
in the  (10) and (01) directions }\\
\end{eqnarray*}

\begin{figure}[h]
\centering                                                                         
\epsfxsize = 4 in \epsfysize = 5in \rotatebox{-90}{\epsfbox{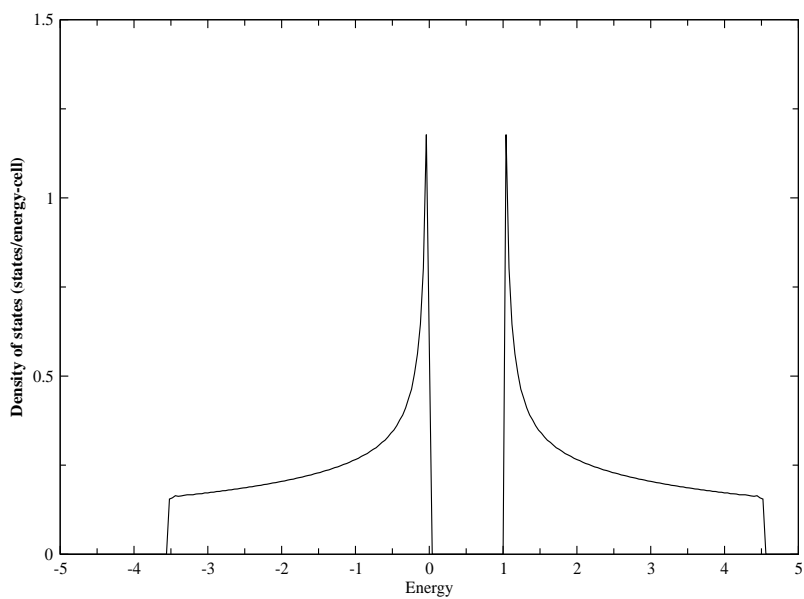}}
\caption{Density of states for the perfectly ordered lattice}
\label{toy0}                                                
\end{figure}

The ordered lattice has a density of states which has a central band gap with 
integrable infinite Van Hove singularities at the two internal band edges, two
kink singularities within the band and the usual square-root singularities at
the external band edges. As expected, the density of states is symmetric about the
band centre ${\mathaccent 22 \epsilon}$ = 0.5. This is shown in figure \ref{toy0}.

\begin{figure}[h]                                                               
\centering                                                                         
\epsfxsize = 6 in \epsfysize = 4.5 in \rotatebox{-90}{\epsfbox{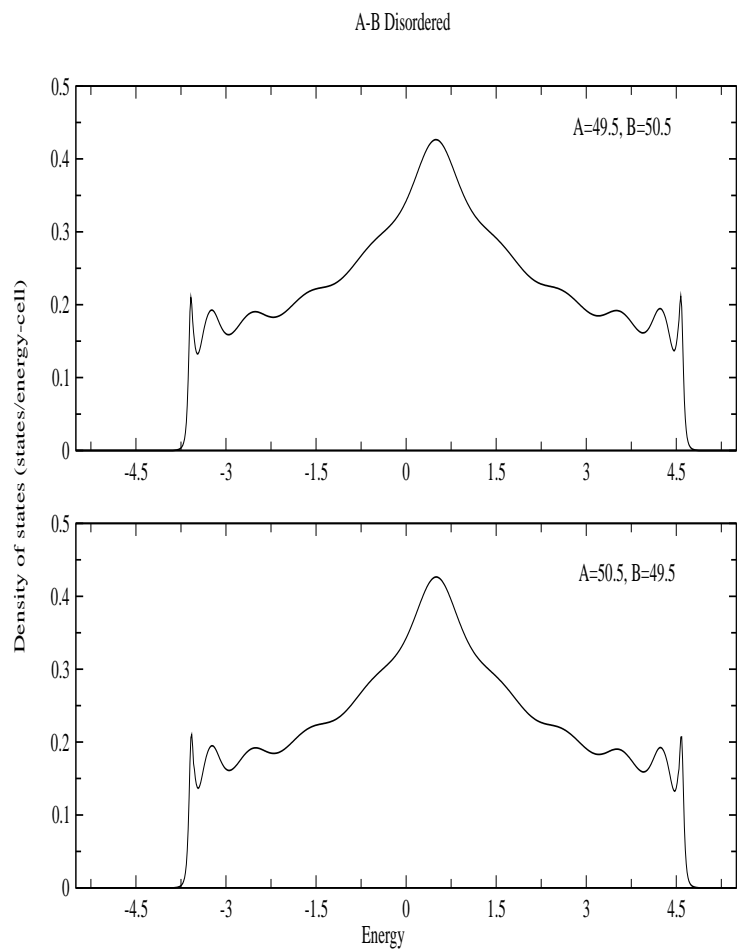}}
\caption{Partial Density of states (states/energy-cell) for the A and B atoms (top,bottom)
for a completely disordered alloy  at
different off-stoichiometric compositions}
\label{toy3}                                                
\end{figure}

The density of states for the perfectly disordered lattice, where each lattice site is 
occupied by either A or B atom with probabilities proportional to their concentrations,
is shown in the figure \ref{toy3}. Disorder washes away the Van Hove singularities and
the central band gap is filled up.  The results are identical to a disordered square 
lattice with lattice vectors ($a/\sqrt{2}$,0) and (0,$a/\sqrt{2}$).

The next figure \ref{toy4} shows the density of states and Fermi energies for the partially
disordered half-filled alloys at just off-stoichiometric compositions (49.5-50.5) and (50.5-49.5). Disorder
washes out the Van Hove singularities, although vestiges of the kink singularities remain. The
signature of the internal infinite singularities show up as peaks, but disorder fills up the
internal band gap.  The band-edge square-root singularities remain as artifacts of the termination procedure. 
Loss of stoichiometry weights the two `bands' differently leading to a loss of symmetry about the band centre.

\begin{figure}                                                               
\centering                                                                         
\epsfxsize = 5.5 in \epsfysize = 5in \rotatebox{-90}{\epsfbox{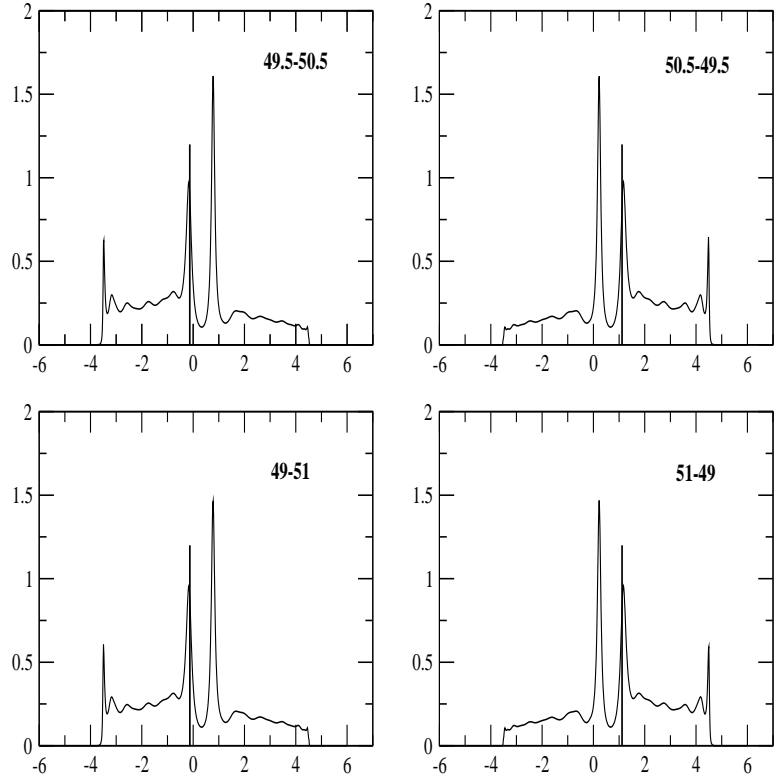}}
\caption{Density of states (states/energy-cell) for partially disordered alloy, at
different off-stoichiometric compositions}
\label{toy4}                                                
\end{figure}                                    

\begin{figure}                                                               
\centering                                                                         
\epsfxsize = 6 in \epsfysize = 5in \rotatebox{-90}{\epsfbox{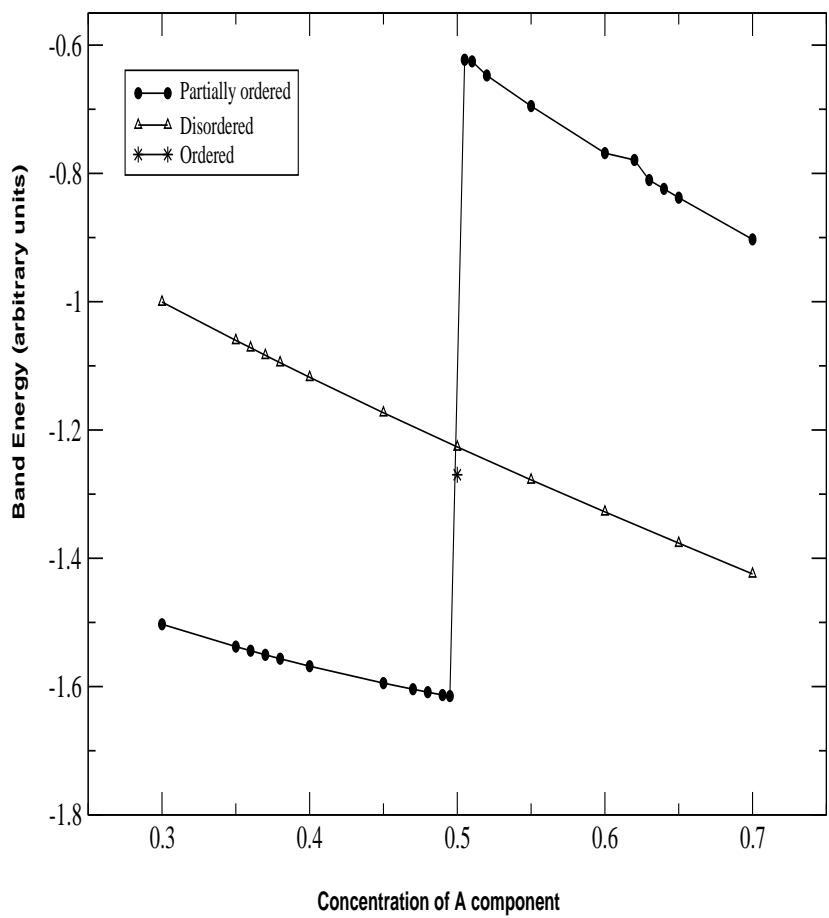}}
\caption{Band energies for the partially ordered, fully ordered and fully disordered alloys
at stoichiometric (50-50) and nearby off-stoichiometric compositions}
\label{toy5}                                                
\end{figure}

The figure \ref{toy5} shows the band energy for the partially ordered alloys. There is a jump at the
stoichiometric composition with the band energy for the ordered alloy falling half way between the
two branches. This jump can be understood by carefully examining the graphs on the left panels and
right panels of the figure \ref{toy5}. The band energy represents the average energy (centre of gravity)
of the portion of the graph to the left of the Fermi-energy. The density of states on the left panel
have greater weight below the Fermi energy as compared to those on the right panel, ensuring the band
energy to be more negative for the compositions sown on the left panels. The sudden jump is the result of
the Fermi energy shifting across the pseudo band-gap as we cross the 50-50 composition and including the
high peak at E=0, which shifts the average energy to higher values. The specific behaviour of the band energy
depends sensitively on the features of the density of states. The specific behaviour for our toy model may
not obtain for realistic alloy systems.

In contrast, the behaviour of the band energy with composition for the fully disordered alloy is smooth,
reflecting the smooth behaviour of the density of states without any internal band gaps. The figure indicates
that for compositions to the left of the 50-50 stoichiometric one, the alloy prefers to be partially ordered.
At 50-50 the ordered alloy is stable for our toy model. While for compositions to the right of 50-50 the
alloy stabilizes in the completely disordered phase.

\section{Remarks and Conclusion}

In this communication we have first proposed a generalization of the ASR for partially ordered binary alloy
systems. Partial ordering of the type where disorder in different sublattices are different has been studied
in particular. We have analyzed the convergence of the recursion technique in the worst scenario case of no disorder and have argued that disorder smoothens out structure in the density of states, so that for disordered
alloys the convergence of the moments is even faster. Finally we have studied a model system, in order to
understand the effects of partial disordering. We are now in a position to apply our formalism to realistic
alloy systems. This will be the subject matter for a subsequent communication.


\newpage
\section*{References}
\begin{thebibliography}{99}
\bibitem{oka1} Andersen O.K.  in  {\sl  Computational  Methods  in  Band
Theory},   eds. P.M. Marcus,  J.F. Janak and A.R. Williams (Plenum,  New York,
1971),  p.178.
\bibitem{oka2} Andersen O.K. and  Jepsen O.,  \PRL {\bf 53} 2571 (1984)
\bibitem{oka3}Andersen O.K., Jepsen O. and Krier G. in {\it Lectures
on Methods of Electronic  Structure  Calculations},   eds.  V.  Kumar,   O.K.
Andersen and A. Mookerjee (World Scientific,  Singapore,  1994),  p.63.
\bibitem{gdm}Ghosh S., Das N., Mookerjee A.,  \JPCM {\bf 9} 10701(1997)
\bibitem{hyd}Haydock R.  \JPA {\bf 7} 2120 (1972) 
\bibitem{hhk}  Haydock R., Heine V. and  Kelly
M. J.,  \JPC {\bf 5} 2845 (1972)
\bibitem{hk}  Haydock R., Heine V. and  
 Kelly M. J.,  {\sl Surf. Sci.} {\bf 38} l39 (1973)
 \bibitem{v35} Haydock R.  {\sl Solid State Physics} (Academic Press, New York)
{\bf 35} (1988)
\bibitem{hay2} Arnold W. T., Haydock R. (private communication) (2001)
\bibitem{kudr} Kudrnovsk\'y., Bose S. K., Anderson O. K.,  \PR {\bf B 43} 4613 (1991)
\bibitem{am1} Mookerjee A.,  \JPC {\bf 6} L205 (1973)
\bibitem{am2} Mookerjee A.,  1973 \JPC {\sf 6} 1340 (1973)
\bibitem{sdm1} Saha T. , Dasgupta I.  and Mookerjee A.,  \JPCM {\bf 6} L245 (1994)
\bibitem{sdm2} Saha T. , Dasgupta I.  and Mookerjee A. ,
 \JPCM  {\bf 8} 1979 (1996)
\bibitem{sm} Saha T. and Mookerjee A., \JPCM {\bf 10} 2179 (1997)
\end{thebibliography}
\end{document}